\documentstyle[eqsecnum,aps,twocolumn]{revtex}

\begin{document}
\draft
\wideabs{
\title
{Susceptibilities of Sr(Cu$_{1-x}$Zn$_x$)$_2$O$_3$ \\
Studied by Quantum Monte Carlo Simulation}

\author
{
Tomohiko {\sc Miyazaki}$^1$, Matthias {\sc Troyer}$^2$,
Masao {\sc Ogata}$^1$ \\
Kazuo {\sc Ueda}$^2$,
and Daijiro {\sc Yoshioka}$^1$
}

\address
{
$^1$Department of Basic Science, Graduate School of Arts and
Sciences\\ University of Tokyo, 3-8-1 Komaba, Tokyo 153, Japan\\
$^2$Institute for Solid State Physics, University of Tokyo,
7-22-1 Roppongi, Tokyo 106, Japan
}

\date
{
June 12, 1997
}
\maketitle

\begin{abstract}
 The effects of non-magnetic impurities randomly doped into a
  two-leg Heisenberg spin ladder are investigated. Using the
  continuous time quantum Monte Carlo loop algorithm we calculate the
  uniform and staggered susceptibilities of such a system.  The obtained
  uniform susceptibility is well described in terms of an effective model of
  weakly interacting local moments induced by non-magnetic impurities
  for a 1\% doping case, but not for higher concentrations.  The
  staggered susceptibility however is significantly enhanced over
  that in the effective model already at 1\% doping.
%  This leads to antiferromagnetic long range order
%  when ladders are coupled weakly.
  Using a mean field approximation for the interladder coupling, we
  explain qualitatively the phase diagram of
  Sr(Cu$_{1-x}$Zn$_x$)$_2$O$_3$.
\end{abstract}
}

%\kword
%{Sr(Cu$_{1-x}$Zn$_x$)$_2$O$_3$, spin ladder, impurity,
%spin-gap, quantum Monte Carlo, N\'eel order
%}
%\begin{document}
%\sloppy
%\maketitle
%%%%%%%%%%%%%%%%%%%%%%%%%%%%%%%%%%%%%%%%%%%%%%%%%%%%%%%%%%%%%%%%%%%%%%%%%%%%%
%------------ 1. INTRODUCTION----------
%------------ 1-1. Spin-gap (\delta=0) --------------------
Spin ladder systems have attracted a lot of theoretical and experimental
interest recently.\cite{review} The physics of these materials is
believed to be closely related to that of planar cuprate
superconductors.
Theoretical studies have shown that
the ground state of the undoped two-leg ladder is a
short range singlet resonating valence bond (RVB) state with a finite
spin-gap $\Delta\approx0.5J$.  This has been
verified experimentally in materials such as SrCu$_2$O$_3$ which has a
gap of about 400K.\cite{Azuma1,Azuma2}
Doping of holes into such a ladder system will probably lead
to superconductivity of preexisting RVB singlet pairs
with d-wave-type symmetries. This seems to be
confirmed by the discovery of superconductivity in the ladder compound
Sr$_{0.4}$Ca$_{13.6}$Cu$_{24}$O$_{41.84}$.\cite{Akimitu}

In this letter we concentrate on another aspect, that of the
suppression of the spin-gap by non-magnetic impurities.
This was first observed in the zinc-doped compound
Sr(Cu$_{1-x}$Zn$_x$)$_2$O$_3$.\cite{AFT,AFT1}
Marked changes in the low temperature behavior occur upon partially
substituting copper with magnetically inert zinc atoms.
Susceptibility and specific heat measurements indicate that the
impurities destroy the spin gap
and lead to a long range N\'eel order at low temperatures.
Doping with 1\% impurities is enough to order the system magnetically
at a N\'eel temperature of 3K.  As the impurity density increases up
to 4\% doping, the transition temperature increases to 8.2K.
Further doping however reduces the N\'eel temperature.
\par
%
%
%------------1.2.2 introduction (review for the other works) ----------
A theoretical model for this material is weakly coupled
spin 1/2 Heisenberg ladders.
Zinc doping leads to a random depletion of this ladder.
Several aspects of an isolated depleted ladder have been
studied.\cite{MKFI,FNST,MDR,SF,Ng,DMRG,NFSF,Iino}
It was shown that a single impurity induces a local
moment with $S=1/2$ around it.\cite{FNST,DMRG}
The effective
interaction between these two induced moments
is ferromagnetic or antiferromagnetic depending on the sublattices
of the two impurities,\cite{SF,DMRG} which does not contradict the
N\'eel order.  The magnitude of the
interaction is determined by the distance between the two impurities, and
decays exponentially.
Based on this localize-moment picture, the enhanced staggered
susceptibility has been discussed,\cite{FNST,Iino}
but the actual magnitude of the staggered susceptibility, N\'eel
temperature and its doping-dependence have not been established.

Our main purpose in this paper is to calculate explicitly the
uniform and staggered susceptibilities and
to discuss the origin of the antiferromagnetic
ordering and the phase diagram.
Using the quantum Monte Carlo loop algorithm,\cite{Loop,Loop1} we obtain
the temperature dependence of susceptibilities, with high precision,
down to extremely low temperatures.
By analyzing these reliable numerical results, we can check whether
the localized-moment picture is valid or not for various values of impurity
density.
It is found that such a picture is valid for only a 1\% doping case.
Then, to determine the phase diagram, we
consider the three-dimensional inter-ladder coupling.
Actually, the interladder coupling inside the plane is frustrated
ferromagnetic\cite{review} and its effect is not clear at present.
Therefore in this paper we assume an unfrustrated antiferromagnetic
coupling which originates from the interplane coupling.
This interaction is assumed to be weak
compared to the intra-ladder couplings and we treat it in a mean field
approximation.  We estimate the staggered susceptibility of the
three-dimensional coupled ladder system as
\begin{equation}
\label{eqn:mft}
 \chi_{\rm MF} = \frac{\chi_{\rm s}}{1 - J' \chi_{\rm s}}.
\end{equation}
Here, $\chi_{\rm s}$ is the staggered susceptibility of an isolated depleted
ladder calculated by quantum Monte Carlo simulation and $J'$ is an effective
inter-ladder coupling.
It is shown that, with this simple mean field approximation,
we can explain the features of the phase diagram.
In particular, we explain why a very small doping
concentration is enough for the N\'eel order and why
overdoping reduces the ordering.

In this study we investigate ladders with randomly-depleted sites.
So far, numerical calculations have been performed for
periodically depleted spin ladders.\cite{MKFI,Iino}
However we will show that the random distribution of impurities is
essential for explaining the phase diagram.
A typical configuration considered here is shown in Fig.~1.
In the randomly-depleted case,
the impurities can cut the ladder into disconnected non-interacting pieces
as shown in the configuration in Fig.~1.
Such configurations have not been studied before.
In our quantum Monte Carlo simulations we have averaged more than twenty
random configurations on a ladder of $1000 \times 2$ sites.
In the following discussion all the error
bars include the statistical error of the Monte Carlo sampling and
that arising from the different configurations of the disorder.
Using this loop algorithm we can achieve Monte Carlo simulations
down to extremely low temperatures such as $T=0.005J$,
which cannot be achieved by other algorithms
Thus we can discuss the temperature dependence precisely.

\par
%
%
%

%------------3. Results-----------------------------------------
First we discuss the uniform susceptibility.
In Fig.~2, the temperature dependence of $\chi T$ is shown for 1\%,
5\%, and 10\% doping.
Initially there is a Curie behavior of free spins at high
temperatures ($T \gg \Delta$), but we are not interested in this
region and it is not shown in Fig.~2.
At intermediate temperatures $T<\Delta$ and for a 1\% doping case,
we observe a plateau with $\chi T = g^2 \mu^2_B x /4k_{\rm B}$, with $x$ being
the impurity density.
The horizontal lines in the figure are the lines with $g^2 \mu^2_B x
/4k_{\rm B}$.
This is consistent with the speculation that a local moment with
$S=1/2$ is induced around an impurity.\cite{FNST}
In this case
the mean distance between the two impurities is so large that the induced
moments are essentially non-interacting in this temperature range.

However at higher levels of doping such as 5\%,
we can see from Fig.~2 that such an
independent localized-moment picture does not hold.
In this case the interaction between the localized moments has already
set in in this temperature range.

Renormalization group arguments\cite{SF} for the random spin chain
predict a limiting value of
$\chi T \rightarrow g^2 \mu^2_B x /12k_{\rm B}$ for $T \rightarrow 0$.
These values are shown by arrows in Fig.~2.
%
%----discussion the result of uniform susceptibility------------
Apparently the predicted zero temperature limit is not observed
for any impurity density down to $T=0.005J$.
Since the localized moment picture seems to hold for 1\% doping,
the present result suggests that the renormalization group argument
becomes valid at temperatures much lower than $0.005J$.
On the other hand, for higher doping cases, the independent
localized-moment picture is no longer valid, so that the
temperature dependence is smeared out.
In this sense, the renormalization group argument does not seem
to be relevant for the experiments.
The results of the uniform susceptibility do
not change within our statistical errors if we restrict the
simulations to connected configurations.
\par
%
%

%------------3.2 Staggered susceptibility -------
%------------3-1 chi(T) -------

The staggered susceptibility (see Fig.~3) exhibits behavior different from
that of the uniform susceptibility.
This is already visible in the
undoped case, where the uniform susceptibility vanishes exponentially
at low temperatures. The staggered susceptibility has a finite value
at zero temperature.\cite{Note}

Impurities lead to a divergent staggered susceptibility at zero
temperature. The magnitude of the staggered susceptibility at a high
impurity density, depends strongly on whether we allow disconnected
configurations or not. If we disallow such configurations we see a
monotonous increase of the staggered susceptibility for a wide range
of dopings.  If we allow disconnected configurations we see that for
more than 10\% doping the staggered susceptibility decreases
again. This explains the drop of the N\'eel temperature at higher
doping levels.
Impurities are needed to induce local moments. Overdoping
however reduces these moments again by dilution.  We will discuss this
effect more quantitatively later in this study.
\par
%------------3-2 chi*T vs T -------
In Fig. 4 we plot the temperature dependence of $\chi_{\rm s} T$.  One can
see immediately that the picture of free spin-1/2 moments induced by
non-magnetic impurities does not describe the staggered susceptibility
well even at 1\% doping.
This behavior, quite different from that of the uniform susceptibility, can
be described as $\chi_{\rm s} T=(a + b/T)T$ at low temperatures.
The constant term $a$ comes from the spins far from the impurities,
and is almost the same as for the undoped case.
The Curie term comes from spins around the impurities.
However, the magnitude
of this term is about 30 times as large as the corresponding term in the
uniform susceptibility.
The magnitude of this term at 1\% doping can be
explained by integrating the distribution of the excess $z$-component of spin
around the impurity site, which is given by the DMRG technique,
with a weakly staggered field.\cite{DMRG}
Thus, the antiferromagnetic correlation between the impurities
cannot be seen in our QMC method down to $T/J=0.005$
at 1\% doping.
The crossover from the Curie behavior of induced spins to
the coupled local moments is estimated to
occur around 4\% doping.\cite{Iino}
\par

Nagaosa {\it et al}.\cite{NFSF} suggested that the staggered susceptibility
diverges as
\begin{equation}
\label{eqn:chi}
 \chi_{\rm s} \propto  \frac{1}{T^{1+2\alpha}},
\end{equation}
with $\alpha \sim 0.22$, based on the renormalization group arguments
of the induced-moment picture.
However Fig.~4 shows clearly that such a divergent behavior does not
exist at $T=0.005J$.
This is for the same reason as the case of uniform
susceptibility as discussed above:
The renormalization group argument
is valid only in temperatures much lower than $0.005J$.

%
%
%
%------Phase diagram-------
We argue here that, even if there is no strong divergence as in eq.\ (2),
the significant enhancement of the staggered susceptibility leads to
antiferromagnetic ordering when the ladders are weakly coupled.
Using the results of staggered
susceptibility and the mean field approximation of eq.(\ref{eqn:mft})
we calculate the N\'eel temperature $T_{\rm N}$ from
\begin{equation}
\label{eqn:Tneq}
 \chi_{\rm s}(T_{\rm N}) = \frac{1}{J'} .
\end{equation}
The exact magnitude of the inter-ladder coupling, $J'$, is not known.
We use the following estimates.
The intra-ladder coupling is taken to
be $J=1000{\rm K}$ as this gives the correct order of magnitude for the
gap in the undoped ladder. The inter-ladder coupling depends on the
density as $J' = J'^0 (1-x)^2$, where $(1-x)^2$ is the
probability of having two spins on adjacent sites.  $J'^0$ is
taken as $0.02J \sim 20$K. This gives the same maximum N\'eel temperature as
seen in the experiments and is consistent with the order of magnitude
of the inter-plane coupling in other cuprate materials.
In Fig.~5, we
show the N\'eel temperatures as a function of impurity density
estimated by this mean field approximation.
As the impurity density increases up to 10\% doping,
the transition temperature increases to 6K.
Further doping however reduces the N\'eel temperature.

%------------4. Discussion-----------------------------------------
%------------4-1 Compare with experimental data--------------------

This phase diagram agrees qualitatively with the experiments on
Sr(Cu$_{1-x}$Zn$_x$)$_2$O$_3$ and can explain both the unexpectedly
high N\'eel temperatures as well as the maximum at substantially low
levels of doping.
The optimal doping ratio is however overestimated as compared to the
experiments. We believe this to be an artifact of the mean field
approximation which takes into account neither the effect of disorder
on the inter-ladder couplings nor the exact structure and
frustration effects of the inter-ladder coupling in the plane.

Note here that we can apply the mean-field approximation
for more than 5\% doping
where there is a correlation between the induced spins.
For small impurity densities (less than 4\%),
the antiferromagnetic correlation between the impurities
cannot be seen in our QMC method at temperatures as low as $T/J=0.005$.
Therefore the validity of applying the mean field approximation
for such small densities
is not clear and the actual N\'eel temperature is
expected to be much lower than the obtained value.
In this sense, the question as to  why
high N\'eel temperatures are observed for small densities
in Sr(Cu$_{1-x}$Zn$_x$)$_2$O$_3$ is unanswered.
\par
%
%
%
%
%
%
%------------------------5.conclusion------------------------------
In summary, we have found that randomly depleted spin ladders with
weak three-dimensional interactions have antiferromagnetic ordering at
low temperatures.
The proposed speculation of independent free spin seems
to be valid only at very low levels of doping of about 1\% but fails at
higher levels of doping.
For higher impurity density, random depletion, which
includes disconnected configurations in the ladder, is important to
suppress the magnetic ordering.
\par
%---------------------------acknowledgments------------------------------
The authors thank M.~Azuma for useful discussions regarding the experimental
results of Sr(Cu$_{1-x}$Zn$_x$)$_2$O$_3$.  The calculations were
performed on the Hitachi SR2201 massively parallel computers at the
University of Tokyo using a parallel C++ Monte Carlo library developed
by one of the authors (M.T.).

%----------------------------Figure Caption-----------------------------
\section*{FIGURE CAPTIONS}
FIG.~1. \quad Example of a randomly depleted ladder. Filled circles denote
copper sites with $S=1/2$ spin and open circles denote $S=0$ impurity
sites. The configuration depicted here includes two impurities on
neighboring spins which break the ladder into disconnected parts.
\par
\vspace{1cm}
FIG.~2. \quad Curie constant per site for several impurity densities
(1\%, 5\%, and 10\%). The horizontal lines and arrows denote the
predictions of the second plateau and the zero-temperature limit
respectively for these doping levels.\cite{SF}
\par
\vspace{1cm}
FIG.~3. \quad Staggered susceptibility per site as a function of
  temperature.  Results for the undoped system and for 1\% and 5\%
  doping are shown by triangles, open circles and closed circles,
  respectively.
  The inset shows results for 1\%, 5\%,
  12.5\%, 15\% and 20\% doping at low temperatures.
\par
\vspace{1cm}
FIG.~4. \quad Temperature dependence of $\chi_{\rm s} T$ for 1\%, 5\% and
10\% impurity doping. Note that the magnitude of the susceptibility
for each impurity density is much larger than that of the uniform
susceptibility.
\par
\vspace{1cm}
FIG.~5. \quad N\'eel temperatures as a function of impurity density as
 estimated in the mean field approximation for the inter-ladder
 coupling.  We have assumed $J=1000$K.
\par

\begin{thebibliography}{99}
\bibitem{Note} This behavior is consistent
with the fact that the ground state is close to the rung-dimer state.
\end{thebibliography}
\end{document}